\documentclass[superscriptaddress,showkeys,showpacs,preprint,pre]{revtex4}

\usepackage{amssymb}
\usepackage{amsmath}
\usepackage{multirow}
\usepackage{hhline}
\usepackage[dvips]{epsfig}
\usepackage{color}

\definecolor{red}{rgb}{1,0,0}

\begin{document}

\title{A simple branching model that reproduces language family and 
language population distributions}
\author{Veit Schw\"ammle}

\affiliation{Laboratoire PMMH, \'Ecole Sup\'erieure de Physique et de Chimie Industrielles, 10 rue Vauquelin, F-75231 Paris, France.}
\affiliation{Centro Brasileiro de Pesquisas F\'{\i}sicas, Rua Xavier Sigaud 150, 22290-180 Rio de Janeiro, Brazil.}

\author{Paulo Murilo Castro de Oliveira}

\affiliation{Laboratoire PMMH, \'Ecole Sup\'erieure de Physique et de Chimie Industrielles, 10 rue Vauquelin, F-75231 Paris, France.}
\affiliation{Instituto de F\'isica, Universidade Federal Fluminense; Av. Litor\^anea s/n, Boa Viagem, Niter\'oi 24210-340, RJ, Brazil.}

\begin{abstract}
Human history leaves fingerprints in human languages. Little is known 
over language evolution and its study is of great importance. Here, we 
construct a simple stochastic model and compare its results to 
statistical data of real languages. The model bases on the recent 
findings that language changes occur independently on the population 
size. We find agreement with the data additionally assuming that 
languages may be distinguished by having at least one among a finite, 
small number of different features. This finite set is used also in 
order to define the distance between two languages, similarly to 
linguistics tradition since Swadesh.

\end{abstract}

\maketitle

\section{INTRODUCTON}
\label{introduction}
The existence of the large number of around $6,000$ languages on Earth 
can be explained through their continuous modification. They come from a 
tree-like evolution being founded from one or few proto-languages 
several thousand years ago. Language evolution is thus a result of the 
particular history followed by humankind and its migration pattern. On 
the other hand, human genetic evolution is also a result of the same 
history. The parallel between language and genetic evolution was 
explored by many researchers since Cavalli-Sforza (for a review, see
\cite{CS1996}), being a precious tool in order to discover unknown 
details of the human past history.

The complex structure of a language changes slowly in time, as for 
instance new words appear and are adopted by a majority of speakers. In 
spite of this complex structure consisting of individually 
difficult--to--predict ``microscopic'' events as for example the 
appearance of a new word, several universal patterns have been reported. 
Statistical data analysis showed universal laws both in actual language 
structure and macroscopic observables as for instance the language 
populations. The histogram of the latter, counting languages according 
to their number of speakers, roughly shows a lognormal 
shape~\cite{Sutherland2003} and will be one of the central points in 
this paper.

In addition to the present situation of nowadays spoken languages, we 
are also interested in their historical course, resulting from the 
branching of ancestor languages. The direct comparison of selected word 
sets of different languages can be used to estimate their historical 
distance, an idea pioneered by Morris Swadesh half a century 
ago~\cite{Swadesh55}. For instance, the measurement of Levenshtein 
distances between two languages gives an idea of the time their first 
common ancestor language existed. The evaluation of the data from this 
analysis showed that historical distances accumulate at a certain 
age independently on the population size~\cite{Petroni2008}, suggesting 
that change occurs with the same rate for all languages~\cite{Wichmann2008}. This result 
challenges the often proposed direct analogy between biological and 
language evolution and demands different approaches. In contrast, the 
global mutation rate of biological species depends on their population 
size, leading for instance to faster changes of the genetic pool in 
smaller populations.

The statistical analysis of highly complex systems like opinion dynamics 
and stock exchange among others showed that their patterns can be 
reproduced by simple agent-based models. The simplification of most of 
the complex low-level mechanisms to random processes has been shown to 
be a valid approach. We follow this idea and will construct a simple 
model that uses the main properties found for the evolution of 
languages. Our stochastic model is based on the following assumptions: 
(i) Languages evolve in a tree-like structure. (ii) The structure is 
modeled by having a probabilistic change rate that does not depend on 
the language's number of speakers. (iii) The space of possible languages 
is finite, the same language can be visited from different evolutionary 
paths. (iv) Each population exponentially increases in time.

Motivated by the increasing popularity of simulating language evolution 
and competition, several analytical and computational models 
concentrated on reproducing the histogram of language sizes. For 
instance, its lognormal-like shape may be explained by simply assuming 
independent language population growth. These independent processes 
naturally lead to a lognormal distribution being result of the central 
limit theorem for multiplied random variables~\cite{Zanette2007}. 
However, the distribution of real languages presents a deviation from 
lognormal behavior for languages spoken by very few people. 
Agent--based models found agreement with the real data furthermore 
reproducing the deviation for small population sizes. This was achieved 
by measuring the histogram in a simulation state before reaching the 
absorbing state~\cite{Stauffer2006b}, or by allowing redundancy in 
creating new languages, i.e. by allowing different historical paths 
leading to the same final language~\cite{Oliveira2007,Oliveira2008}. The 
latter model also considers the gradual geographical conquest of new 
territories, introduced in~\cite{Viviane2006a,Viviane2006b}, and until 
now may be seen as the most appropriate one as it not only reproduces 
the histogram of language sizes but also the distribution of languages 
family sizes recently reported~\cite{Wichmann2005}. Another recent work 
proposes that the histogram of language sizes deviates from the 
lognormal shape also for large populations displaying power-law decays 
at both extremes~\cite{Schwaemmle2008d}. For a thorough review of 
language models see refs.~\cite{Schulze2008,Wichmann2007}. We try here 
to gather the principal findings from previous models in order to 
construct a yet simpler model with a reduced number of parameters that 
reproduces the real data sets.

This work is organized as follows. The following section introduces the 
model. The next section presents our results and their direct comparison 
to real data. Finally, we conclude and discuss our findings in the last 
section.

\section{Branching model}
Our model is agent--based, i.e. each language has its own characteristic 
traits. A language is characterized by its number of speakers, given by 
a real number, its structure condensed into a bit-string of $L$ 
features, and an integer number defining the language family it makes 
part of. Time evolution takes place in discrete time steps having the 
following rules applied to each language present at a given time step. 
(i) The number of speakers is multiplied with the factor $1+G$ leading 
to its exponential increase. (ii) The current bit-string characterizing 
the language has one of its bits flipped at a random position, from 0 to 
1 or vice-versa. (iii) The grown language divides into two with 
probability $b$. In the case of branching, three additional processes 
occur. First, the population is distributed by giving a randomly chosen 
part between $0\%$ and $50\%$ of the ancestor one to the new branching 
language and leaving the remaining population to the ancestor one. 
Afterwards, the bit-string of the new language suffers a flip of 
one of its bits, compared with the ancestor, making both therefore 
differ from each other by one feature. Third, if the new language still 
belongs to the first family, it founds a new language family with 
probability $F=0.5$. Otherwise it obtains the family label from the 
ancestor one.

The simulation initializes with one single language having all bits set 
to $0$. As there are no extinctions, the total number of branches 
increases exponentially with time until we stop the simulation after 
reaching the previously given threshold of $N_\text{L,max}$ branches. We 
could easily consider also language extinctions with a given probability, but 
this would introduce a further, unnecessary parameter into the model: 
the absence of some branches simply corresponds to a smaller value of 
the branching parameter.

\begin{figure}[tb]
 \centering
 \includegraphics[angle=0,width=0.8\textwidth]{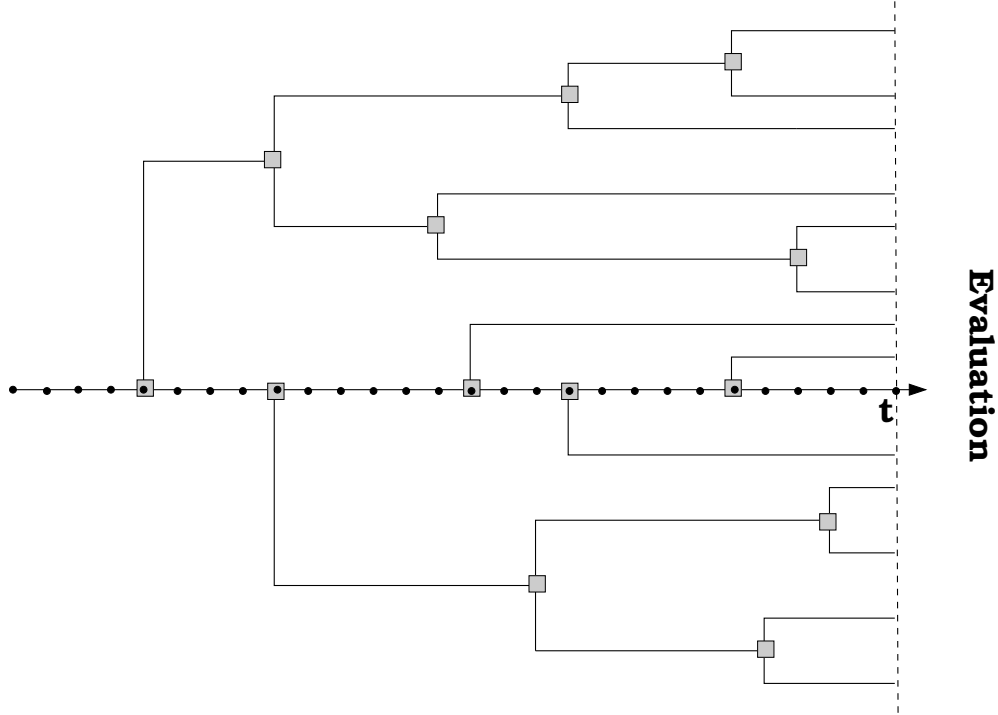}
\caption{Schematic description of the language model. Mutations occur
 according to the same rate for all languages, at the time steps 
 visualized by small black dots along the time axis. At each time step, 
 a branching event occurs with probability $b$, visualized by grey 
 squares. After reaching a previously defined number of branches, 
 vertical line on the right, the simulation stops and the data are evaluated.}
\label{fig:1}
\end{figure}

After the final time step we process the data, consisting of a large 
number of branches, each having its corresponding bit-string, its number 
of speakers and the family it pertains. The data evaluation defines 
which branches differ, and which are bundled to represent only one 
language. More precisely, we compare the bit-strings of each pair. If 
the language structures are identical, both populations are added to 
form a unified language. Additionally, if their family labels, assigned 
during the previous dynamic evolution, differ from each other, the 
family of the one with the larger population is kept. Depending on the 
number of language features, i.e. bit-string length $L$, the number of 
languages after evaluation will be less or equal $2^L$. Fig.~\ref{fig:1} 
shows a scheme of our model.

\section{Comparing the results to real data}
As a classification of languages, we take the database of the ethnologue 
which is accessible through the internet~\cite{Grimes2000}. There, they
also provide the number of speakers of a language, defined by the number 
of people speaking a language as their mother tongue. This data will be 
compared to the simulational results.

Both real and simulational data present ranges of population sizes over 
about nine decades. Therefore it is convenient to plot the histogram in 
the following way: We count the number of languages in a bin with 
population sizes between $2^n$ and $2^{n+1}$. As a consequence, the size 
of the bins increases exponentially. We plot these numbers without 
dividing them through the length of the bin. Therefore, they do 
\emph{not} correspond to the frequency of languages with population 
sizes within the given ranges.

\begin{figure}[tb]
 \centering
 \includegraphics[angle=270,width=0.8\textwidth]{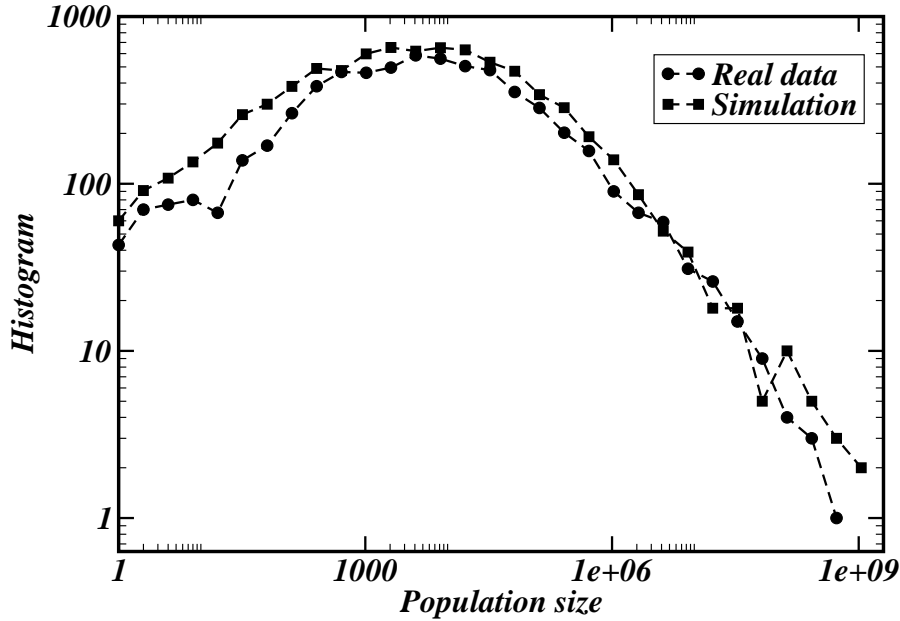}
\caption{Comparing the histograms of language populations: real data in 
 the world $versus$ simulations. The vertical axis counts the number of 
 languages spoken by the number of people displayed along the horizontal 
 axis.}
\label{fig:2}
\end{figure}

By making a histogram of language population sizes (Fig.~\ref{fig:2}), 
we found good agreement between simulational results and real data for a 
bit-string length of $L=13$, a branching probability $b=0.01$, and a 
growth rate $G=0.023$. The simulation stopped as soon as the tree had 
more than $40,000$ different branches, corresponding to $1,024$ time 
steps and a total population of $1.3 \cdot 10^{10}$ speakers. As we 
defined the languages to be different only if they exhibit distinct 
features, the final number of languages was $8,125$, near the maximum 
value allowed with $L=13$, indicating the possibility of re-visiting the 
same language by different historical paths. On the other hand, within 
similar numbers of languages but with larger values of $L$, the 
possibility of re-visiting the same language more than once becomes 
negligible. In this case, our simulations also result in lognormal-like
distributions as in Fig.~\ref{fig:2}, but with a stronger deviation observed at 
the very left part, for languages spoken by very few people. Based on 
these observations, we can tentatively interpret these deviations: 
languages spoken by very few people, separated in small groups which 
nevertheless exchange some experiences with each other, may be the 
results of different historical paths leading to the same final 
language, i.e. historical redundancy. This multiple-historical path for 
the same language indeed occurs also for large languages (spoken by 
millions of people). For instance, different regions of the same large 
country have little language differences, which nevertheless were not 
captured by the linguistics finite-set of words adopted in order to 
distinguish one language from another. 
This leads to the decrease of counted languages with small populations.
According to this interpretation, the quoted 
deviations would become more accentuated if one is able to enlarge the set of words 
adopted nowadays in order to distinguish languages, i.e. by enhancing 
the ``resolution''. Yet according to this interpretation, the resolution 
currently adopted by professional linguists is just enough to 
distinguish the main real human languages from each other. The measuring 
instrument is designed with the resolution degree one needs for the 
object at hands, no more.

The simulational results strongly depend on the random seed set at the 
beginning of the simulation in order to generate the sequence of 
pseudo-random numbers. The randomly chosen time steps at which the first 
branching events occur are crucial, and can lead to different 
simulational results. The same behavior is also expected in reality, 
the real historical path depends on contingencies occurred during the 
evolution, and this dependence is supposed to be stronger for 
contingencies occurred at the very beginning. Unfortunately, one is not 
able to re-run the real evolution again, only computer simulations like 
the current one allow this repetition, by taking different random seeds. 
By doing so, the curve of the histogram of population sizes moves 
towards larger or smaller populations. Thus different random seeds 
result in a horizontal shift on the histogram. By changing the total 
number of time steps we are additionally able to perform vertical 
shifts. However, in any case the lognormal shape of the distribution, 
including the deviation for languages spoken by few people remains the 
same. Histograms with smaller widths and smaller heights could be 
interpreted as describing early stages of the human languages evolution.

The two parameters, branching probability $b$ and growth rate $G$, have 
a similar effect. By increasing the branching probability, the number of 
time steps needed to reach the previously fixed maximum number of 
branches is smaller, and therefore the final total population becomes 
smaller as it grows exponentially with time. The shape of the histogram 
remains unaltered. Nevertheless, the strong dependence on the random 
seed makes a scaling of this parameter difficult. On the other side, the 
growth rate has the opposite effect on the simulational results. By 
linearly increasing it, the total population with fixed number of branches at the end of the simulations
increases exponentially.

\begin{figure}[tb]
 \centering
 \includegraphics[angle=270,width=0.8\textwidth]{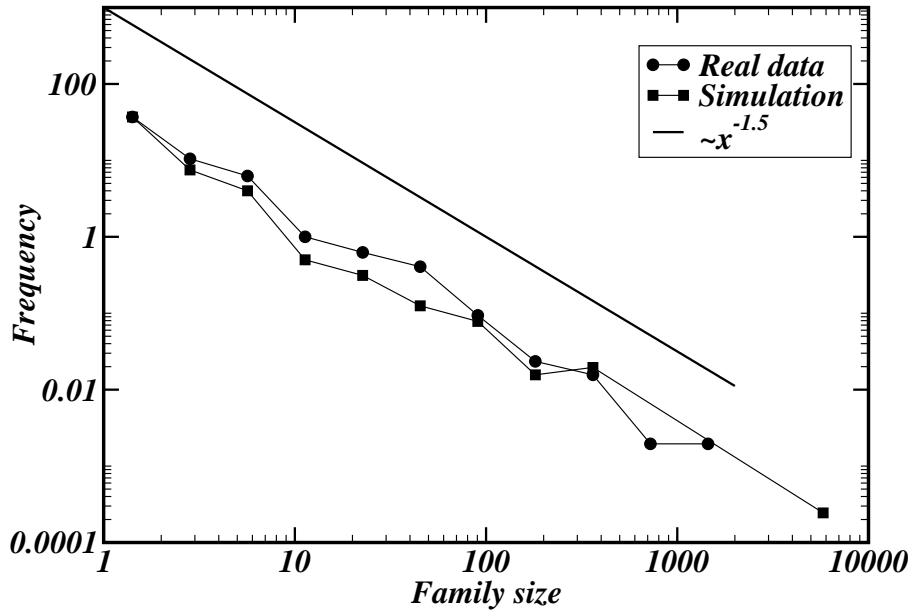}
\caption{Frequency of language families comparing simulational and real 
 data. The same power-law is found in both systems.}
\label{fig:3}
\end{figure}

Now we look if our model also reproduces the other simple macroscopic 
law that has recently been reported for 
languages~\cite{Wichmann2005}. The size distribution of language  
families displays a power-law decay. The size of a family is obtained by 
counting its languages. In Fig.~\ref{fig:3}, we show the frequency of 
the families, again for a binning over powers of 2. We count the number 
of families with sizes between $2^n$ and $2^{n+1}$ and divide the 
obtained number by the length of the binning, $2^n$. The simulational 
data and the real one agree very well with each other, exhibiting a 
power-law with an exponent of about $-1.5$. The parameter values are the 
same as the ones for the simulation shown in Fig~\ref{fig:2}.

\section{Conclusions}

Based on a few basic assumptions, we constructed a simple branching 
model and compared its statistical results with real data of human 
languages and families of languages. The good agreement shows that our 
model may be considered as a minimal model for the evolution of 
languages. In particular, we did not need to consider any geographical 
issue.

The only important assumptions we need are: (i) All languages descend 
from a single, ancient mother-tongue; (ii) All languages change 
according to the same and constant rate, independent of the number of 
speakers; (iii) Bifurcation events (a single language leading to two 
different ones) occur according to a smaller rate, also constant and the 
same for all languages; (iv) The different features allowing to 
distinguish two languages from each other can be condensed into a 
finite, small set; (v) The founder of a new language family always 
belongs to the original family founded by mother-tongue. The last point 
does not mean one cannot go further into the taxonomic classification, 
by taking into account language genera, etc. We simply did not look to 
this subject within this work.

\bibliographystyle{unsrt}
\bibliography{language}

\end{document}